\documentclass{emulateapj}

\usepackage{graphicx}
\usepackage{epstopdf}
\usepackage{hyperref}
\usepackage{url}
\usepackage{fnpos}

\newcommand{\gray}{$\gamma$-ray\ } \newcommand{\grays}{$\gamma$-rays\ }

\shorttitle{Extragalactic Gamma Ray Background}
\shortauthors{Stecker and Venters}

\begin{document}
\DeclareGraphicsExtensions{.pdf,.gif,.jpg}

\shorttitle{EGB Components}
\shortauthors{Stecker \& Venters}

\title{Components of the Extragalactic Gamma Ray Background}

\author{Floyd W. Stecker\altaffilmark{1} and Tonia M. Venters\altaffilmark{2}}
\altaffiltext{1}{Astrophysics Science Division, NASA Goddard Space Flight Center, Greenbelt, MD 20771}
\altaffiltext{2}{NASA Postdoctoral Program Fellow, Goddard Space Flight Center, Greenbelt, MD 20771}

\begin{abstract}
We present new theoretical estimates of the relative contributions of unresolved blazars and star-forming galaxies to the extragalactic \gray background (EGB) and discuss constraints on the contributions from alternative mechanisms such as dark matter annihilation and truly diffuse \gray production. We find that the {\it Fermi} source count data do not rule out a scenario in which the EGB is dominated by emission from unresolved blazars, though unresolved star-forming galaxies may also contribute significantly to the background, within order-of-magnitude uncertainties. In addition, we find that the spectrum of the unresolved star-forming galaxy contribution cannot explain the EGB spectrum found by EGRET at energies between $50$ and $200$ MeV, whereas the spectrum of unresolved FSRQs, when accounting for the energy-dependent effects of source confusion, could be consistent with the combined spectrum of the low-energy EGRET EGB measurements and the {\it Fermi}-LAT EGB measurements.
\end{abstract}


\section{Introduction}

Studies of the extragalactic \gray background (EGB) can provide insight into high energy processes in the universe and, as such, has been the subject of much debate, particularly concerning the roles of extragalactic astrophysical sources and new physics. Recent data from the Large Area Telescope (LAT)\footnote{Hereafter we shall refer to the {\it Fermi}-LAT instrument simply as {\it Fermi}.}  on board the {\it Fermi Gamma Ray Space Telescope} allow for a reassessment of the possible astrophysical origins of the EGB, which could improve our understanding of \gray production in these objects and provide more robust constraints on the more exotic scenarios. However, in order to determine the strength and spectrum of this isotropic background one needs to proceed from the raw photon count data by determining to the best extent possible the detector sensitivity, the intrinsic events produced by the larger charged particle flux impinging on the detector, and the much larger \gray foreground within our Galaxy resulting from cosmic-ray interactions with photons and gas nuclei. Such analyses have been made for both the Energetic \gray Experiment Telescope (EGRET) aboard the {\it Compton Gamma Ray Observatory} \citep{sre98,str04}  and {\it Fermi} \citep{lat10}. 

Various extragalactic \gray production scenarios have been explored theoretically as candidate components that could contribute significantly to the observed background. Among those considered are the unresolved astronomical sources, such as active galactic nuclei (AGN) \citep{pad93,ste93,sal94,chi95,ste96,kaz97,chi98,muk99,muc00,gio06,nt06,der07,pv08,ino09,ven09, ven10,aba10}, star-forming galaxies \citep{pav02, fie10, mak10}, and starburst galaxies \citep{tho07, ste06, mak10}. The large majority of associated extragalactic sources thus far detected by both EGRET and {\it Fermi} are blazars \citep{har99,1cat}, {\it i.e.}, those AGN for which the jet is closely aligned
with the observer's line-of-sight \citep{bla79}, including \gray loud flat spectrum radio quasars (FSRQs) and BL Lacertae-type objects. It is expected that since blazars comprise the largest class of identified extragalactic \gray sources, unresolved blazars should contribute significantly to the EGB. Additionally, just as our Galaxy produces $\gamma$-rays, it is expected that \grays are produced in other galaxies, and as such, unresolved galaxies might also contribute to the EGB with the most significant contribution originating from the population of actively star-forming galaxies \citep{ste75, pav01,pav02,fie10,mak10}. Interesting truly diffuse mechanisms that could contribute to the EGB involve cosmic ray interactions with intergalactic gas and the cosmic background radiation \citep{faz66,ste73,dar07,kes03} and electromagnetic cascades produced by interactions of very high and ultrahigh energy particles with the extragalactic background light \citep{kal09, ber10, ahl10,ven10}, as well as more exotic scenarios such as dark matter annihilation~\citep{sil84, ste85, rud88, ste89, ste89a, rud91,ull02} and decay \citep{oli85,ste86, iba08}\footnote{For reviews on dark matter annihilation, see \citet{jun96} and \citet{ber05}.}.

In this paper, we estimate the contributions to the EGB from unresolved extragalactic \gray sources of various types and compare them with the EGB obtained from analysis of {\it Fermi} data. In doing so, we also take into consideration the effects of both the completeness of the {\it Fermi} flux limited blazar survey and the important effect of source confusion owing to the energy dependent angular resolution of the {\it Fermi}-LAT detector. We will then briefly discuss the implications of possible truly diffuse emission mechanisms to the EGB.

\section{EGRET and Fermi Resolved Sources and the \gray Log N - log S relation}\label{sec:datalogNlogS}

\begin{figure}[t]
\begin{center}
\resizebox{2.5in}{!}
{\includegraphics[trim = 0 0 0 1.5mm, clip]{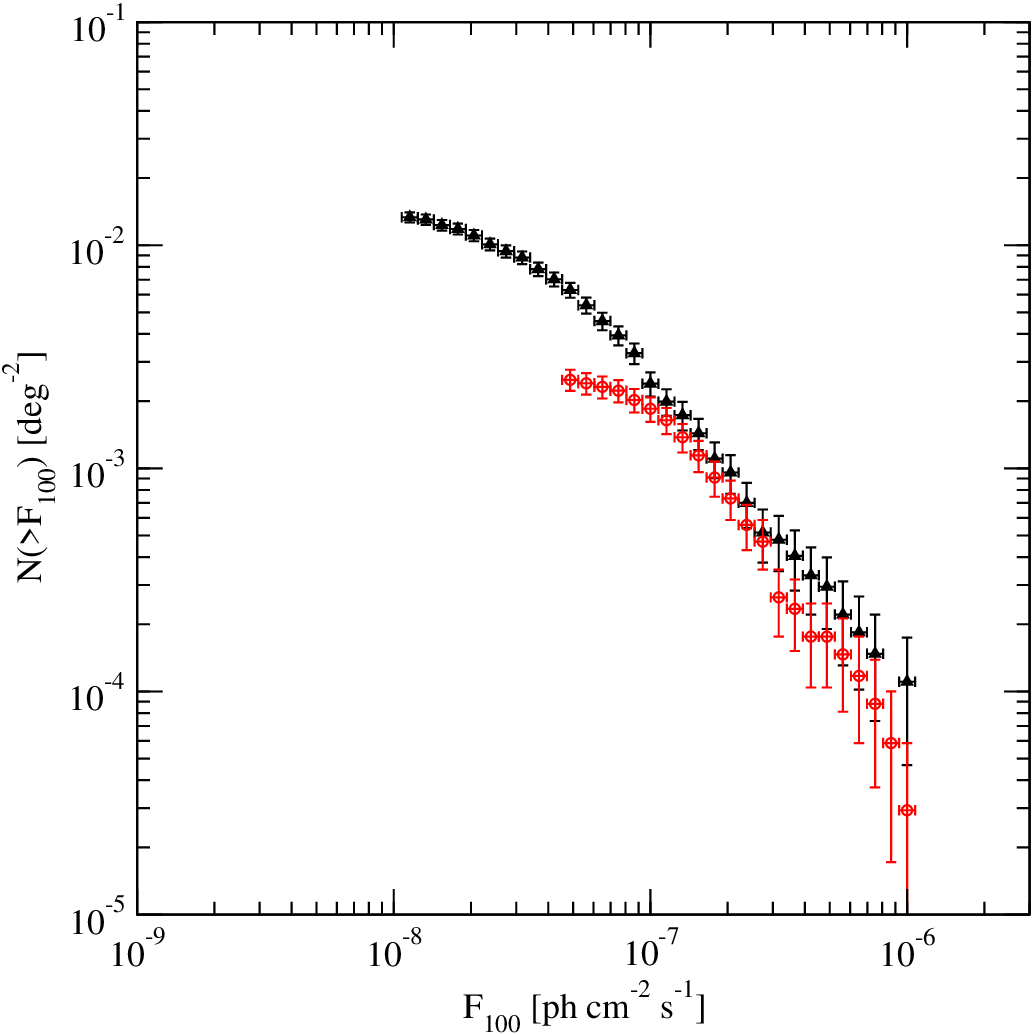}}
\caption{Source count distributions for {\it Fermi} \citep[black triangles;][]{1cat} and EGRET \citep[red circles;][]{rei01} blazars.
\label{fig:logN/logSdata}}
\end{center}
\end{figure}

In Figure \ref{fig:logN/logSdata} we plot the number of blazars observed per square degree {\it versus} blazar flux integrated above $100$ MeV for both EGRET \citep{rei01} and {\it Fermi} \citep{1cat}. In the case of {\it Fermi}, the source spectra were extrapolated from a power-law fit above a fiducial energy of $1$ GeV \citep{pop10}. The offset between the resolved source count data obtained by the two detectors is a result of differences in their sensitivity calibrations\citep{tho93,1cat}. The fluxes of the {\it Fermi} sources are reported to be systematically brighter than those reported by EGRET. 

We note that the 50\% completeness in the {\it Fermi} survey is estimated to be reached at an effective flux limit of $\sim 2 \times 10^{-8}$ cm$^{-2}$ s$^{-1}$ \citep{pop10}. The flux limit varies with galactic latitude and source hardness. Nevertheless, the effective flux limit indicated by both the turnover in Figure \ref{fig:logN/logSdata} and the Monte Carlo simulation in \citet{pop10} is $\sim 2 \times 10^{-8}$ cm$^{-2}$ s$^{-1}$. By assuming that the {\it Fermi} survey is 100\% complete at $\sim 7 \times 10^{-8}$ cm$^{-2}$ s$^{-1}$ \citep{pop10} and comparing the data for EGRET and the {\it Fermi} shown in Figure \ref{fig:logN/logSdata}, we estimate that the EGRET survey is 50\% complete at $\sim 8 \times 10^{-8}$ cm$^{-2}$ s$^{-1}$.

One may note that the flux level attained for 50\% completeness for the {\it Fermi} extragalactic blazar survey is only about four times fainter than that for the EGRET survey even though the {\it Fermi}-LAT is sensitive to sources $\sim 30$ times fainter than that of EGRET. It is important to note that there are several factors that determine the efficiency of a \gray telescope for detecting extragalactic sources, among which are:  (1) the flux of the source, (2) the spectral index of the source, (3) the intrinsic detector background from cosmic-ray induced events, (4) the foreground from the Milky Way, and (5) the diffuse extragalactic background. The {\it Fermi}-LAT was designed to reach its optimal effective area for \grays with energies near and above $1$ GeV, whereas  EGRET was designed to reach its optimal effective area for \grays with energies near and above $100$ MeV. As such, {\it Fermi} is more sensitive to sources with hard spectral indices, particularly for the faintest sources observable by {\it Fermi}. Thus, it is difficult to make a direct comparison between EGRET and {\it Fermi}. However, we note that the positions of the turnovers in the data presented in Figure \ref{fig:logN/logSdata} provide a good indication of the observational situation.

\section{Determination of the EGB from Point Sources}\label{sec:formalism}

In general, the total contribution of a given population of sources to the EGB is found by convolving the source spectrum, $F_{\rm ph}(E_0,z,L_{\gamma})$, measured at energy, $E_0$, of one source of \gray luminosity, $L_{\gamma}$, at a given redshift, $z$, with the comoving number density of sources at that redshift per luminosity interval, $n(L_{\gamma},z)=d^2N/dL_{\gamma}dV_{\rm com}$. We then integrate over the comoving volume, $V_{\rm com}$, and \gray luminosity:
\begin{widetext}
\begin{equation}\label{eqn:totalcontgeneral}
I_E(E_0) = \int_0^{z_{\rm max}} \int_{L_{\gamma,{\rm min}}}^{L_{\gamma,{\rm max}}} F_{\rm ph}(E_0,z,L_{\gamma})n(L_{\gamma},z)\frac{d^2V_{\rm com}}{dzd\Omega}dL_{\gamma}dz\,,
\end{equation}
\end{widetext}
where $I_E(E_0)$ is the EGB intensity given in units of photons ${\rm cm^{-2} \, \, s^{-1} \, \, sr^{-1} \, \, GeV^{-1}}$, $L_{\gamma,{\rm max}}$ depends on the redshift, source spectrum, and the detector sensitivity, and we have differentiated the comoving volume with respect to redshift and solid angle, $\Omega$. However, as the different types of likely contributing \gray sources have distinct spectral characteristics and redshift and luminosity distributions, we must consider each case separately.

Note that in Equation \ref{eqn:totalcontgeneral} we have neglected the effect of \gray absorption due to pair-production interactions with the extragalactic UV photons.  The inclusion of \gray absorption would result in a steepening in the collective point source spectrum at the high-energy end \citep{sal98,ven09}, though a possible contribution from electromagnetic cascade photons might mitigate the steepening at higher energies \citep{ven10}.  However, recent {\it Fermi} constraints on the \gray opacity imply that the UV background is likely to be fairly low \citep{opa10}, and as such, the absorption and the resulting cascades will only have a small effect on the EGB.

\subsection{The Contribution to the EGB from Unresolved Blazars}\label{subsec:blazarform}

In determining the blazar contribution to the EGB, we follow the procedure outlined in \citet{ven09}.  We approximate the blazar \gray spectrum as a power law in energy defined by the \emph{photon} spectral index, $\Gamma$ ($F_{\rm ph} \propto E^{-\Gamma}$).  The spectral indices of the population of blazars form a distribution with some spread \citep[the spectral index distribution, SID;][]{ste96,vp07}; hence, the number density of blazars is defined as
\begin{equation}
n(L_{\gamma},z,\Gamma) = \rho_{\gamma}(L_{\gamma},z)p_{L}(\Gamma)= \frac{d^3N}{dL_{\gamma}dV_{\rm com}d\Gamma}\,,
\end{equation}
where $\rho_{\gamma}(L_{\gamma},z) = d^2N/dL_{\gamma}dV_{\rm com}$ is the blazar \gray luminosity function (GLF) giving the comoving number density of blazars per luminosity interval and $p_{L}(\Gamma) = dN/d\Gamma$ is the normalized blazar SID accounting for spectral bias (see Section \ref{sec:inputs}). $L_{\gamma}$ is the \gray luminosity at the fiducial energy, $E_f$ (taken to be $100$ MeV), defined as $E_f^2$ times the differential photon luminosity, $L_{\rm ph} = d^2N_{\gamma}/dtdE$ measured at $E_f$.  $L_{\gamma}$ is related to the integral flux greater than $E_f$, $F( > E_f)$, by
\begin{equation}\label{eqn:lgamma}
L_{\gamma} = 4\pi D^2 (\Gamma - 1) (1+z)^\Gamma E_f F( > E_f)\,,
\end{equation}
where $D$ is the distance measure for the Friedman-Robertson-Walker cosmology\footnote{We take $H_0 = 70 \mbox{ km s}^{-1} \mbox{ Mpc}^{-1}$, $\Omega_m = 0.3$, $\Omega_{\Lambda} = 0.7$, and $\Omega_{r} \ll 1$.}:
\begin{equation}
D = \frac{c}{H_0} \int_0^z \left[\Omega_\Lambda+\Omega_m(1+z')^3\right]^{-1/2}\,dz'\,.
\end{equation}
A given blazar of \gray luminosity, $L_{\gamma}$, at a redshift, $z$, with a \gray photon spectral index, $\Gamma$, has a measured photon flux of 
\begin{equation}
F_{\rm ph} (E_0, z, L_\gamma, \Gamma) = \frac{L_\gamma}{4 \pi E_f^2
  [d_L(z)]^2}(1+z)^{2-\Gamma}\left(\frac{E_0}{E_f}\right)^{-\Gamma}\,,
\end{equation}
where $d_L(z) = D(1+z)$ is the luminosity distance. Thus, the total contribution to the EGB at a given energy, $E_0$, from unresolved blazars (the collective unresolved blazar intensity) is determined by integrating the contribution from each individual blazar fainter than the detector sensitivity, 
\begin{widetext}
\begin{equation}\label{eqn:totalcontblazars}
I^{\rm bl}_{E}(E_0) = \int_{-\infty}^{\infty} \int_0^{z_{\rm max}} \int_{L_{\gamma,{\rm min}}}^{L_{\gamma,{\rm max}}} F_{\rm ph}(E_0,z,L_{\gamma},\Gamma)\rho_{\gamma}p_L(\Gamma)\frac{d^2V_{\rm com}}{dzd\Omega}\,dL_{\gamma}\,dz\,d\Gamma\,,
\end{equation}
\end{widetext}
where $z_{\rm max} = 5.0$ and $L_{\gamma,{\rm max}}$ is determined from Equation \ref{eqn:lgamma} taking $F( > E_f) = F_{\rm min}$, with $F_{\rm min}$ being the minimum flux capable of being resolved by {\it Fermi}. We have also integrated Equation \ref{eqn:totalcontgeneral} over the blazar spectral index. 

\subsection{The Contribution to the EGB from Unresolved Star-forming Galaxies}\label{subsec:galaxyform}

By applying the same procedure that we use to determine the background from unresolved blazars (see Section \ref{subsec:blazarform}), we calculate the contribution from unresolved star-forming galaxies by determining the \gray photon flux as a function of energy for one galaxy and then convolving with and integrating over the appropriate cosmological distributions.  Is is expected that, as in the Milky Way, the \gray emission for a star-forming galaxy comes mainly from the decay of $\pi^{0}$ mesons produced by cosmic-ray interactions with interstellar gas~\citep{ste70}. The resulting \gray production spectrum has been calculated by many authors 
\citep{ste70,ste73,ste79,cav71,step81,der86,mor97,str04a,str07,str10,kel06,kam06,mor09}.  For our calculation we adopt the $\pi^0$ emissivity given by~\citet{ste79} renormalized upwards by 25$\%$ to be consistent with the local emissivity measured by {\it Fermi}~\cite{gd09}. We note that there is also emission arising from electron bremsstrahlung, but the contribution is likely to be small, particularly above $100$ MeV \citep{gd09}. We also neglect the emission from Compton interactions that may contribute significantly to galaxy spectra above 10 GeV, particularly for starburst galaxies \citep{ste77, hun97, mw09, str10}. We also note that the cosmic ray spectrum should, in fact, vary from galaxy to galaxy since it depends on energy dependent leakage and each galaxy has a different morphology and magnetic field configuration. This uncertainty can affect the predicted slope of the EGB spectrum at high energies, but it does not affect the absolute value of the predicted background at $\sim 200$ MeV.

The \gray photon luminosity is related to the \gray production spectrum per hydrogen atom, $q_{\rm H}(E)$, by
\begin{equation}
L_{\rm ph}(E) = \left<q_{\rm H}(E)\right>N_{\rm H}\,,
\end{equation}
where $\left<q_{\rm H}(E)\right>$ is found by averaging $q_{\rm H}(E)$, the differential \gray production spectrum per hydrogen atom, over the galaxy, and $N_{\rm H}$ is the number of hydrogen atoms in the galaxy in the form of both atomic (HI) and molecular (${\rm H_2}$) hydrogen.  Thus, the \gray photon flux for a galaxy at redshift $z$ as observed at energy $E_0$ is 
\begin{equation}
F_{\rm ph}(E_0,z) = \frac{1}{4 \pi D^2}\left<q_{\rm H}[E_0(1+z)]\right>N_{\rm H}(z)\,.
\end{equation}
The rate of production of \grays from $\pi^{0}$ decay is proportional to the flux of cosmic rays, which we assume to be proportional to the supernova rate. The supernova rate is expected to be proportional to the rate of formation of higher mass stars, which is, in turn, proportional to the
overall star formation rate assuming a universal initial mass function. 

Assuming then that the rate of production of \grays from $\pi^{0}$ decay is proportional to the star formation rate (SFR) for a galaxy, we can relate the $\left<q_{\rm H}\right>$ of the galaxy to that of the Milky Way, $\left<q^{\rm MW}_{\rm H}\right>$:
\begin{equation}
\frac{\left<q_{\rm H}\right>}{\left<q^{\rm MW}_{\rm H}\right>} = \frac{\Psi(z)}{\Psi(z=0)}\,,
\end{equation}
where $\Psi$ is the SFR of the galaxy, and we take $\left<q^{\rm MW}_{\rm H}\right>$ to be some fraction, $f_q$, of the locally measured\footnote{That is, since $q^{\rm MW}_{\rm H}$ is calculated in the literature assuming the cosmic ray flux as measured in the solar neighborhood.} $q^{\rm MW}_{\rm H}$. We determine $f_q$ by integrating the radial profile of the flux of cosmic rays weighted by $r^2$. Using the radial profiles calculated by \citet{ste77b}, we obtain $f_q \sim 0.825$. Thus, the galaxy spectrum becomes
\begin{equation}\label{eqn:galfluxgeneral}
F_{\rm ph}(E_0,z) = \frac{1}{4 \pi D^2}f_q q^{\rm MW}_{\rm H}[E_0(1+z)]\frac{\Psi(z)}{\Psi(0)}N_{\rm H}(z)\,.
\end{equation}

The calculation of the amount of gas in a galaxy is subject to a considerable degree of uncertainty, especially at high redshifts (see Section \ref{subsec:sfginput}).  As such, rather than focusing on one particular model, we calculate the star-forming galaxy contribution for three different models arising from different sets of assumptions.  In so doing, we seek to explore various possibilities and highlight the uncertainty.

\subsubsection{Galaxy Contribution Determined from the Schecter Function and An Evolving Gas Fraction}\label{subsubsection:schecter}

One method for determining the number of hydrogen atoms in a galaxy is to assume that the mass of gas in the galaxy is some fraction of its stellar mass:
\begin{equation}\label{eqn:nhpap}
N_{\rm H}(z) = \frac{f_{\rm gas}(z)}{1-f_{\rm gas}(z)}\frac{M_{\ast}}{m_{\rm H}}\,,
\end{equation}
where $f_{\rm gas}(z) = M_{\rm gas}/(M_{\rm gas}+M_{\ast}) \propto (1+z)^{0.9}$ is the gas fraction given in \citet{pap10}, and we have neglected the possible contribution of helium to the gas mass of a galaxy. Substituting Equation (\ref{eqn:nhpap}) into Equation (\ref{eqn:galfluxgeneral}) and approximating $\Psi(z)/\Psi(0) \sim \dot{\rho}_{\rm SFR}(z)/\dot{\rho}_{\rm SFR}(0)$, we get:
\begin{equation}
F_{\rm ph}(E_0,z,M_{\ast}) = \frac{f_q q^{\rm MW}_{\rm H}[E_0(1+z)]}{4\pi m_{\rm H}D^2 }\frac{\dot{\rho}_{\rm SFR}(z)}{\dot{\rho}_{\rm SFR}(0)}\frac{f_{\rm gas}(z)}{1-f_{\rm gas}(z)}M_{\ast}\,,
\end{equation}
where $\dot{\rho}_{\rm SFR}(z)$ is the cosmic SFR (CSFR) given by $\log(\dot{\rho}_{\rm SFR}(z)) = -2.06 +3.39\log(1+z)$ for $z < 1.3$ and $\dot{\rho}_{\rm SFR}(z) \sim \mbox{const.}$ for $1.3 \leq z \leq 4.0$ \citep{ly10}.

To get the total contribution to the EGB, we convolve with the comoving number density of star-forming galaxies per stellar mass interval as a function of redshift and integrate:
\begin{equation}\label{eqn:totalcontsfgalsch}
I^{\rm gal}_E(E_0) = \int_{0}^{z_{\rm max}} \!\!\!\!\! \!\!\! dz\, \frac{d^2V_{\rm com}}{dzd\Omega} \!\! \int_{M'_{\rm min}}^{M'_{\rm max}} \!\!\!\!\! \!\!\!\! dM' F_{\rm ph}(E_0,z,M')\Phi(z,M')\,,
\end{equation}
where $\Phi(z,M') = d^2N/dM'dV_{\rm com}$ is the Schecter function for stellar mass with parameters as determined in \citet{els08}, $M'$ is given by $M_{\ast} = 10^{M'}M_{\odot}$, and we take $M'_{\rm min} = 8.0$ and $M'_{\rm max} = 12.0$.

\subsubsection{Galaxy Contribution Determined from IR Luminosity Functions}\label{subsubsec:infrared}

Alternatively, we can determine the \gray spectrum of a galaxy by assuming that the \gray luminosity of the galaxy is proportional to some power of its SFR \citep{fie10, mak10,M31lat10}: $L_{\rm ph} \propto \Psi^{\alpha}$. Since $L_{\rm ph}(E) = \left<q_{\rm H}(E)\right>N_{\rm H}$ and $\left<q_{\rm H}(E)\right> \propto \Psi$ (as demonstrated in Section \ref{subsubsection:schecter}), 
\begin{equation}
N_{\rm H} = \left(\frac{A\Psi_{\rm MW}}{\int \left<q^{\rm MW}_{\rm H}(E)\right>dE}\right)\Psi^{\alpha-1}\,,
\end{equation}
where $A$ and $\alpha$ are the best-fit parameters of the above power law determined from {\it Fermi} observations of star-forming galaxies in the Local Group and their SFRs (see Section \ref{subsubsec:inputIR}).  Assuming the \citet{cha03} initial mass function, the SFR of a galaxy is related to its total infrared luminosity, $L_{\rm IR}$,
\begin{equation}\label{eqn:IRSFRconv}
L_{\rm IR} = 1.1 \times 10^{10}L_{\odot}\left(\frac{\Psi}{M_{\odot}{\rm \,\, yr^{-1}}}\right)
\end{equation}
\citep{hop10}. The \gray photon flux for a galaxy at redshift $z$ is given by
\begin{widetext}
\begin{equation}
F_{\rm ph} (E_0,z,L_{\rm IR}) = \frac{1}{4\pi D^2} \left(\frac{A}{\int q^{\rm MW}_{\rm H} dE}\right)\left(\frac{L_{\rm IR}}{1.1\times 10^{10}L_{\odot}}\right)^{\alpha}q^{\rm MW}_{\rm H}[E_0(1+z)].
\end{equation}
\end{widetext}
Then, the total contribution to the EGB is found by convolving the galaxy photon flux with an infrared luminosity function, $\Phi(z,L_{\rm IR}) = d^2N/dL_{\rm IR}dV_{\rm com}$ and integrating over infrared luminosity and redshift:
\begin{equation}\label{eqn:totalcontsfgalir}
I^{\rm gal}_E(E_0) = \int_{0}^{z_{\rm max}} \!\!\!\!\! \!\!\! dz\, \frac{d^2V_{\rm com}}{dzd\Omega} \!\! \int_{L_{\rm IR,min}}^{L_{\rm IR,max}} \!\!\!\!\! \!\!\!\! \!\!\!\!\! dL_{\rm IR} F_{\rm ph}(E_0,z,L_{\rm IR})\Phi(z,L_{\rm IR})\,,
\end{equation}
where we take $L_{\rm IR,min} = 10^{10}L_{\odot}$ and $L_{\rm IR,max} = 10^{15}L_{\odot}$.

\subsubsection{Galaxy Contribution Determined from the Cosmic Star-formation Rate and the Star-formation Efficiency}\label{subsubsec:csfr}

Another alternative is to relate the cosmic density of hydrogen in star-forming galaxies to the cosmic star formation rate.  Given that stars are formed in giant molecular clouds (GMCs), it is reasonable to assume
\begin{equation}
\dot{\rho}_{\rm SFR} \sim \xi({\rm H_2})\rho_{\rm H_2}\,,
\end{equation}
where $\dot{\rho}_{\rm SFR}$ is the cosmic star formation rate density (see Section \ref{subsubsection:schecter}), $\rho_{\rm H_2}$ is the cosmic molecular hydrogen density in star-forming galaxies, and $\xi({\rm H_2})$ is the star formation ``efficiency" (SFE) of molecular hydrogen \citep{big08,gne09,bau10}. \citet{ler08} measure the SFE to be $\sim (5.25 \pm 2.5)\times 10^{-10} {\rm yr^{-1}}$ and to be roughly constant over a wide range of conditions\footnote{More precisely, $\xi({\rm H_2}) \sim \epsilon/\tau_{\rm ff}$, where $\epsilon$ is the percentage of gas involved in forming stars and $\tau_{\rm ff} \propto \rho^{-1/2}$ is the \emph{local} free-fall timescale of the gas.  However, using this relation requires knowledge of the local density of the gas and a better understanding of the formation of GMCs than presently exists (see Section \ref{subsec:sfginput}). We also note that the measurements obtained by \citet{ler08} were taken from a sample of low-redshift galaxies.  The SFE could actually evolve with redshift \citep{bau10}.}. We can relate the density of atomic hydrogen to the density of molecular hydrogen through the average mass ratio of atomic and molecular hydrogen ($\mathcal{R} = \left<M_{\rm HI}/M_{\rm H_2}\right>$) in star-forming galaxies, $\rho_{\rm HI} \sim \mathcal{R}\rho_{\rm H_2}$.  The average mass ratio of atomic and molecular hydrogen can be found by integrating radial profiles of the gas surface densities of star-forming galaxies found in \citet{ler08}, resulting in $\mathcal{R}\sim 0.9$.  Note that in so doing, we only integrate the profiles out to the optical radius since recent surveys indicate that star formation is extremely inefficient beyond this radius \citep{big10}\footnote{We should note that even though star formation is extremely inefficient beyond the isophotal radius, there is still gas beyond this radius. However based on the radial profiles determined in \citet{ste77b}, we do not expect cosmic rays to propagate much beyond the isophotal radius; hence, we expect gamma-ray production beyond this radius to be low.}. Thus, with appropriate modifications to Equation \ref{eqn:galfluxgeneral}, we find the \gray flux from a particular redshift 
\begin{widetext}
\begin{equation}
F_{\rm ph}(E_0,z) =  \frac{f_q(1+\mathcal{R})}{4\pi m_{\rm H}\xi({\rm H_2})D^2}q^{\rm MW}_{\rm H}[E_0(1+z)]\frac{\dot{\rho}^2_{\rm SFR}(z)}{\dot{\rho}_{\rm SFR}(0)} \frac{dV_{\rm com}}{dz}dz\,.
\end{equation}
\end{widetext}
Differentiating with respect to solid angle $\Omega$ and integrating over redshift results in an equation for the total contribution to the EGB:
\begin{widetext}
\begin{equation}
I^{\rm gal}_E(E_0) = \frac{f_q(1+\mathcal{R})}{4\pi m_{\rm H}\xi({\rm H_2})\dot{\rho}_{\rm SFR}(0)} \int_{0}^{z_{\rm max}}  \!\!\!\! \frac{1}{D^2}q^{\rm MW}_{\rm H}[E_0(1+z)]\dot{\rho}^2_{\rm SFR}(z) \frac{d^2V_{\rm com}}{d\Omega dz} dz\,.
\end{equation}
\end{widetext}

\section{Observational Inputs and Considerations}\label{sec:inputs}

\subsection{\gray Blazars}\label{subsec:inputblazars}

Of the nearly $1500$ resolved point sources observed by {\it Fermi} in the first year, $573$ are associated with blazars (First {\it Fermi} catalog (1FGL); \citealt{1cat}).  Thus, blazars comprise the largest class of astrophysical objects associated with \gray sources.  Naturally, unresolved blazars have long been suspected of providing, at least, a substantial contribution to the EGB, though the exact amount remains in debate and depends on various assumptions as to constructing GLFs and redshift distributions~\citep {pad93,ste93,sal94,chi95,ste96,kaz97,chi98,sre98,muk99,muc00,gio06,nt06,der07,kne08,pv08,ino09,ven09,pop10,ven10}. A detailed discussion of all of the assumptions that go into these calculations is beyond the scope of this paper (though, for a detailed discussion of the \citealt{chi98} calculation, see \citealt{gamma01}), and it is likely that many of these models will be updated in light of {\it Fermi} data. However, we note that as discussed in \citet{pop10}, the source counts predicted by \citet{der07} and \citet{muc00} fall short of the {\it Fermi}  observations of resolved sources above $5 \times 10^{-8} \, {\rm ph} \, {\rm cm}^{-2} \, {\rm s}^{-1}$. In calculating the blazar contribution to the EGB, we assume functional forms for the blazar GLF and SID and fit them to 1FGL data, accounting for errors in measurement of blazar spectral indices and the spectral bias inherent in a flux-limited catalog.

\subsubsection{Source Counts for Faint Unresolved Blazars: Theory Meets Observations}\label{subsubsec:blazarsc}

Determining the GLF from observations relies on the ability to associate \gray blazars with lower energy counterparts for which redshifts can be measured \citep[for discussion, see][]{ven09,ven10}.  However, making the necessary association can be complicated by the angular resolution of the {\it Fermi}-LAT, which is limited by electron scattering in the LAT detector and is much poorer than that of more traditional telescopes (see Figure \ref{fig:angres}). The resulting wide point-spread function results in significant source confusion at energies below $\sim$ 1 GeV even for fluxes well above the {\it Fermi}-LAT sensitivity.  In principle, one could construct source counts from fluxes integrated above an energy for which source confusion is less of a hinderance, but doing so would limit the already suppressed blazar number statistics.  Thus, rather than construct a luminosity function solely from \gray blazars with redshifts, we employ the \citet{ste96}\footnote{See also \citet{nt06}.} approach of determining the luminosity function from wavebands with larger samples and smaller positional error circles.

As in \citet{ste96}, we take the functional form of the FSRQ luminosity function from radio observations \citep{dun90}, but corrected for the present cosmological parameters. The \gray luminosity of a blazar is then determined from its radio luminosity.  The average correlation between the radio and \gray luminosities of blazars is determined by fitting the bright end of modeled source counts\footnote{In doing so, we also include the blazar SID (see Section \ref{subsubsec:blazarSID}).} to that of the observed \gray source counts ($\chi^2_{\rm reduced} \sim 0.4$).  In so doing, we find that \gray luminosity integrated from $100$ MeV to $100$ GeV is $\sim 10^{3.2}$ times $\nu L_{\nu}$ in radio, in agreement with the results obtained using recent {\it Fermi} observations\footnote{One might be concerned that the effect of the new radio-$\gamma$ correlation would be to increase the blazar background with respect to the \citet{ste96} model. However, we note that \citet{ste96} distinguished between ``quiescent'' and ``flaring'' blazars and used separate radio-$\gamma$ correlations (and spectral properties) for each subpopulation. In this paper, we make no such distinction, so a comparison between our results and those of the \citet{ste96} model is not straightforward. Using the \citet{ste96} radio-$\gamma$ correlation for quiescent blazars from \citet{ste96} would yield a smaller blazar background, but it would also under-predict the bright end of the \gray source counts. To compensate, one would have to add a flaring component (as per \citet{ste96}) to fit the data, which would also contribute to the background. In effect, such a procedure is equivalent to our method of fitting the radio-$\gamma$ correlation to the data.} \citep{gir10, lat10,ghi10a,mah10}. We should also note that we include sources out to $z \sim 5$, but as demonstrated in \citet{ven09}, the emission is dominated by sources with $z \lesssim 2$ (consistent with expectations from the redshift distribution predicted by the GLF). As such, we do not expect this choice to significantly impact the results.

\begin{figure}[t]
\begin{center}
\resizebox{2.5in}{!}
{\includegraphics[trim = 0 0 0 1mm, clip]{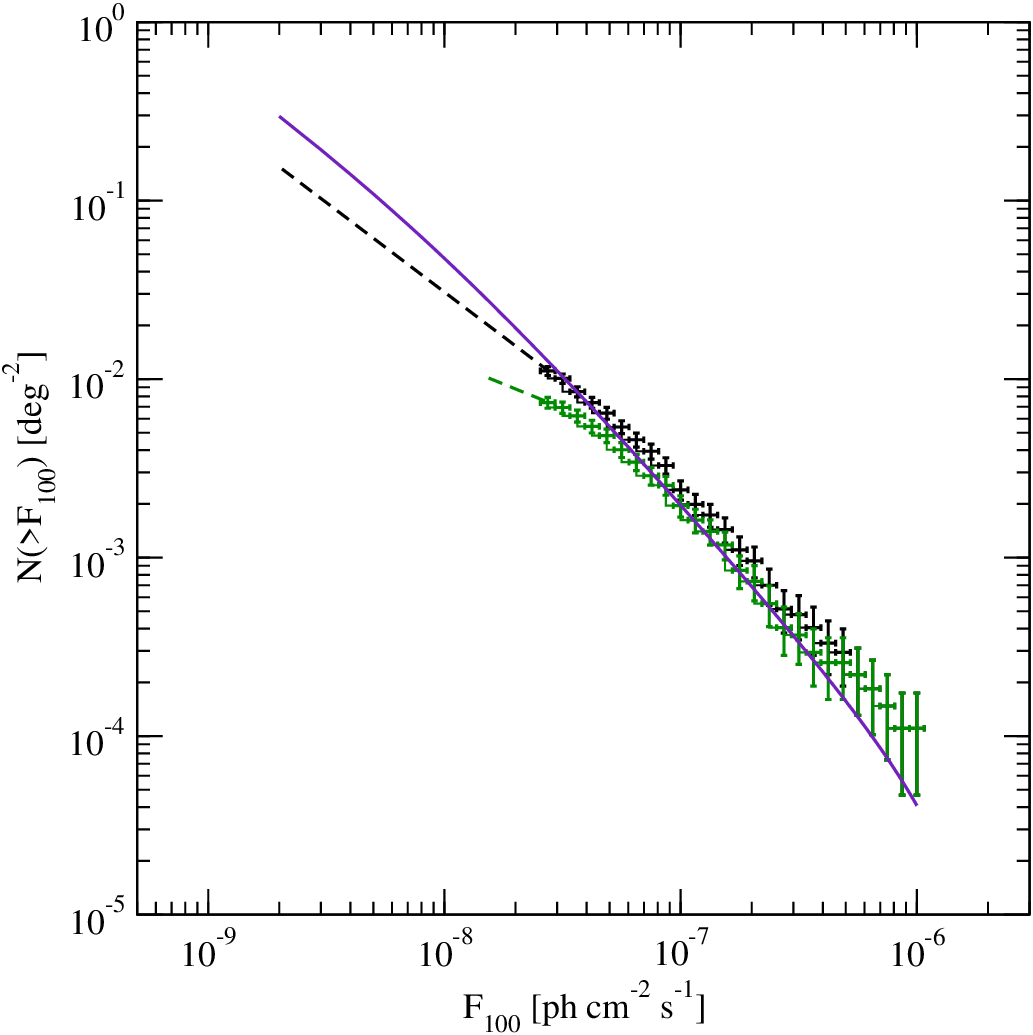}}
\caption{Bright end of the source count distributions for all blazars (including BL Lacs; black data points) and FSRQs (light green data points).  Our model fit to the data is shown by the solid (purple) line.  The dashed lines are the faint-end slopes determined in \citet{pop10} by including a modeled Monte Carlo {\it Fermi}-LAT efficiency.\label{fig:logN/logStheory}}
\end{center}
\end{figure}

The resulting modeled source count distribution for FSRQs (solid) is presented in Figure \ref{fig:logN/logStheory} along with the distributions of the brighter FSRQs resolved by {\it Fermi} (light data points) and all blazars (dark data points). We also show the forms of the
unresolved source count distributions obtained from a Monte Carlo modeled {\it Fermi}-LAT sensitivity calculation performed by \citet{pop10} as dashed lines.  All of the models separate from the data at fainter fluxes as the source counts fall off very rapidly due to the survey incompleteness and source confusion.  Thus, the determination of the \emph{true} faint-end shape directly from the source counts can be severely hindered. 

In an effort to mitigate the effect of survey completeness, \citet{pop10} modeled the {\it Fermi}-LAT efficiency from a Monte Carlo simulation and then divided the differential source counts by this efficiency.  Thus, the calculated source counts for faint sources strongly depend on such modeled efficiencies. Indeed at fluxes near the {\it Fermi}-LAT sensitivity limit, the efficiency is extremely small and model dependent; hence, in flux bins with low number statistics, the source counts are multiplied by a very large and uncertain number.  The result of such a procedure is that even though one source is not statistically different from two sources, whether one source is seen or two could result in different modeled counts for faint sources. It has also been argued that unassociated \gray sources are likely to be dominated by known classes of \gray sources, especially blazars \citep{mir10} and will have a contribution to the EGB even though they will not have been included in the source count distributions.  It is also important to note that in determining the blazar contribution to the EGB from measured source count distributions, source confusion could not be taken into account since its effect depends on the source density (see Section \ref{subsubsec:sourceconf}), which is exactly the unknown quantity that the observer seeks to determine. As such, it is likely that the blazar contribution to the EGB will be underestimated in analyses based solely on the measured source count distributions. Since in this paper we use a theoretically determined source count distribution, our model gives a source count density from which one can determine the effect of source confusion (see Section \ref{subsubsec:sourceconf}). Such differences between our analysis and that given in \citet{pop10} result in different calculations of the unresolved blazar contribution to the EGB:  \citet{pop10} conclude that blazars can only account for less than 25\% of the EGB\footnote{In Section 8 of \citet{pop10} a value of 40\% of the EGB is obtained if one extrapolates to zero flux.}, while our analysis indicates that blazars could possibly account for the bulk of the EGB. 

The $F_{100}$ fluxes included in the source count distributions presented in Figure \ref{fig:logN/logSdata} are not actually \emph{measured} $F_{100}$ fluxes since the 1FGL catalog does not include $F_{100}$ fluxes.  In order to determine the $F_{100}$ fluxes, we make use of Equation 1 of \citet{pop10} to \emph{extrapolate} $F_{100}$ fluxes from measured differential fluxes determined at the pivot energies. \citet{1cat} define the pivot energy as that energy for which the differential flux is minimal. We note that for 1FGL blazars, the average of the pivot energies is $\sim 1$ GeV. Thus, the source count distributions might be more representative of source counts for sources brighter than the {\it Fermi} sensitivity above $1$ GeV rather than $100$ MeV. In effect (and also as the result of spectral bias; see Section \ref{subsubsec:blazarSID}), the  source counts could underestimate the number of blazars with $F_{100}$ fluxes above the {\it Fermi} sensitivity, mostly impacting the faint end of the source count distributions. As such, analyses on unresolved blazars based solely on source counts likely underestimate the blazar contribution to the EGB. This effect could provide an explanation for the fact that {\it Fermi} observes roughly as many BL Lacs as FSRQs even though BL Lacs are intrinsically fainter than FSRQs.  Since the \gray spectra of BL Lacs are harder than those of FSRQs they are more easily observed by {\it Fermi} than FSRQs.

\subsubsection{The {\it Fermi}-LAT and EGRET Angular Resolutions and Source Confusion}\label{subsubsec:sourceconf}

As mentioned previously, the large angular resolutions of pair-production \gray detectors such as EGRET and the {\it Fermi}-LAT\footnote{\url{http://www-glast.slac.stanford.edu/software/IS/glast_}\ \url{lat_performance.htm} .} result in significant source confusion, particularly for faint sources and for energies below $\sim 1$ GeV. Since the angular resolution for the {\it Fermi}-LAT is similar to that of EGRET at $\sim 100-200$ MeV, the capability of {\it Fermi} to resolve faint sources is similar to that of EGRET at these energies. Hence, if the EGB does indeed consist of unresolved sources, then the EGB measurements of the detectors should be similar at $\sim 100-200$ MeV, whereas at $\sim 1$ GeV, the improved angular resolution of {\it Fermi} with respect to that of EGRET should result in a lower measurement of the EGB by {\it Fermi} than that of EGRET owing to the enhanced ability of {\it Fermi} to resolve point sources  \citep{ste99}.

\begin{figure}[t]
\begin{center}
\resizebox{2.5in}{!}
{\includegraphics[trim = 0 0 0 1mm, clip]{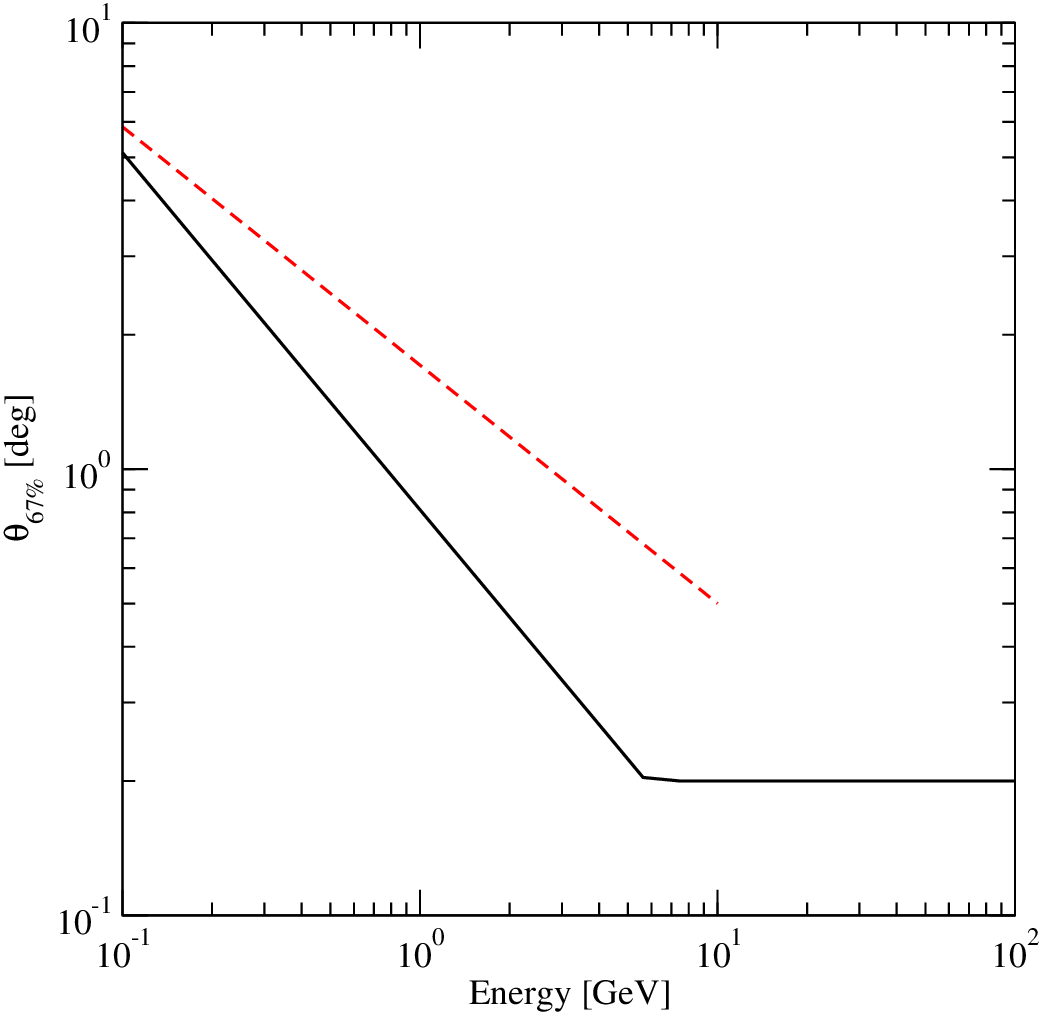}}
\caption{The angular resolution as a function of energy for the {\it Fermi}-LAT \citep[solid black line;][]{atw09} and EGRET \citep[dashed red line;][]{tho93}.
\label{fig:angres}}
\end{center}
\end{figure}

The probability, $\cal{P}$, of finding a nearest neighboring source with $S \ge S_{\rm lim}$ within the minimum angular separation, $\theta_{\rm min}$, for a source density, $N$, is given by 
\begin{equation}\label{eqn:prob}
{\cal{P}} (\le \theta_{\rm min}(E)) = 1 - \exp(-\pi N\theta^2_{\rm min}(E)).
\end{equation}  
For our source confusion criterion, we take the acceptable probability limit, ${\cal{P}}_{\rm min} \sim 0.1$, and $\theta_{\rm min}(E) \sim \theta_{67\%}(E)$ approximately given by
\begin{equation}
\theta_{67\%}(E) = 5.12^{\circ} \times \left(\frac{E}{100 \, {\rm MeV}}\right)^{-0.8}
\end{equation}
\citep{atw09} (see Figure \ref{fig:angres}).  Then, the source density criterion (SDC) is found by inverting Equation \ref{eqn:prob}:
\begin{equation}
N_{\rm SDC} = - \frac{\ln (1-{\cal{P}}_{\rm min})}{\pi \theta^2_{\rm min}(E)}\,.
\end{equation}
The limiting source flux, $S_{\rm SDC}$, is then determined from the modeled source counts.

If one were to think of the {\it Fermi}-LAT as an ordinary telescope with $\theta_{\rm min} = \theta_{67\%}$ where $\theta_{67\%}$ is the half angle for a beam the contains $67\%$ of the photons. The source density criterion would correspond to $\sim 1/10$ sources per beam. For sources with $F_{100} \la 1 \times 10^{-7} \, \mbox{ph cm}^{-2} \mbox{ s}^{-1}$, the probability for finding another \gray source of similar or greater flux within the error circle is quite high; hence, many sources that, in principle, should be resolvable are, in fact, \emph{unresolved}. At fluxes close to the sensitivity limit, this probability is so high that faint sources are \emph{indistinguishable} and will contribute to the measured EGB of the detector in question (for the {\it Fermi}-LAT resolution at $100$ MeV, the source criterion corresponds to $N_{\rm SDC} \sim 1.3\times 10^{-3} \, {\rm sources}/{\rm deg}^{2}$; for reference, at $F \sim 2 \times 10^{-9} \, {\rm ph}\, {\rm cm}^{-2} \, {\rm s}^{-1}$, our model predicts a much larger source density of $\sim 0.3 \, {\rm sources}/{\rm deg}^{2}$). Thus, the effect of source confusion compounded with the separate effect of detector sensitivity is to flatten the faint end of an observationally derived source count distribution. 

We should emphasize that this definition of source confusion is not the same as that employed by \citet{pop10}. Source confusion as discussed in this section refers to the probability that for a given source density, a source has a nearest neighbor within the angular resolution of the detector with a flux greater than or equal to the flux limit.  In \citet{pop10}, the term is applied to a \emph{detected} source associated with a given \emph{real} source for which the measured flux of the detected source is greater than the flux of the real source (plus three standard deviations) by a given amount (the analysis was performed on simulated data to determine the impact on the actual data). In effect, the former definition identifies the limit at which the source density is sufficiently large so that individual sources \emph{cannot} be resolved, and only fluctuations are observed. The latter definition applies to the probability that a given detected source is in fact the superposition of several sources. However, the \citet{pop10} criterion is based on the assumption that a source can be \emph{resolved}. As such, the source would have to be significantly brighter than the background, including the sources within the error circle of the detector. In the case of many sources with similar fluxes, none of the sources would be resolved. In the case of one bright source and several fainter sources, the flux of the detected source would be dominated by the flux of the brightest source and likely would escape the \citet{pop10}. Thus, under this criterion, it is not surprising that they conclude that there is very little source confusion. A common treatment of source confusion in measured source counts is a fluctuation analysis, but as yet, such an analysis has not be performed on {\it Fermi} source counts.

\subsubsection{Blazar Spectra}\label{subsubsec:blazarSID}

The spectral indices of the population of blazars form a distribution with a given finite spread \citep{ste96,vp07}, resulting in curvature in the spectrum of the collective intensity of unresolved blazars due to the increasing relative importance at high energies of blazars with harder spectral indices \citep{ste96,pv08}. Thus, the determination of the blazar SID is crucial in determining the correct spectrum of the unresolved blazar contribution.

In determining the blazar SID from survey data, one must carefully account for uncertainties in measurement of the spectral indices.  In so doing, we follow the likelihood method of \citet{vp07}, fitting {\it Fermi} 1FGL FSRQs to a gaussian SID.  We determined the maximum-likelihood gaussian SID parameters (mean, $\Gamma_0$, and spread, $\sigma_0$) to be $\Gamma_0 = 2.45$ and $\sigma_0 = 0.15$.

We must also account for the effect of spectral bias of the 1FGL catalog. Even in a flux-limited catalog, low-luminosity, high-redshift blazars that would be most likely to appear are those with spectral indices that are harder than most of the population.  The 1FGL catalog is not actually flux limited (see Section \ref{sec:datalogNlogS}), rather it consists of sources above a given threshold in test statistic\footnote{For 1FGL, this threshold is $25$, corresponding to a statistical significance of $\sim 4\sigma$.}, which depends on source spectra and the background.  However, as demonstrated in \citet{ven10a}, applying the likelihood analysis to the sample of FSRQs with fluxes $\ga 7\times 10^{-8} \, {\rm ph \, cm^{-2} \, s^{-1}}$ and galactic latitudes $\ga 10^{\circ}$ \citep[as per][]{pop10} does not appreciably change the SID parameters ($\Gamma_0 = 2.45$, $\sigma_0 = 0.16$)\footnote{See also \citet{pop10}.}.  Thus, even though the 1FGL catalog is not flux limited, we can assume that the sample of 1FGL FSRQs is approximately flux limited.  We can then follow the method of \citet{ven09} in correcting for the sample bias inherent in a flux-limited catalog. In doing so, we apply a correction factor, $\hat{M}(\alpha)$, to the SID 
\citep[for derivation, see][]{ven09}:
\begin{equation}
p_{L}(\alpha) = \frac{\hat{p}(\alpha)}{\hat{M}(\alpha)}\,,
\end{equation}
where $\hat{p}(\alpha)$ is the SID corrected for measurement uncertainty in the spectral indices, and
\begin{equation}
\hat{M}(\alpha) \propto \int_{F_{\gamma,\rm min}}^{\infty} dF_\gamma
\frac{1}{F_\gamma}\int_{z=0}^{\infty} dz \hat{\rho}_\gamma (\alpha,z,F_\gamma)
\frac{dV_{\rm com}}{dz}(z)\,.
\end{equation}

\subsubsection{Summary of Differences with the \citet{ste96} Blazar Model}

To summarize, our approach in calculating the blazar contribution to the EGB is similar to that of \citet{ste96} with some notable differences:
\begin{enumerate}
\item The model has been updated to make use of the current cosmological parameters. The change in cosmology has little impact on the results.
\item The model considers only FSRQs, and flaring blazars are not considered separately from quiescent blazars. (For possible impact, see item 3.)
\item The $\gamma$-ray--radio relation has been updated to be consistent with multi-wavelength observations of FSRQs conducted by {\it Fermi} and radio telescopes ($L_{\gamma} \sim 10^{3.2} \times \nu L_{\nu}$ as opposed to $10^{2.6}$, see Section \ref{subsubsec:blazarsc}). As noted in Section \ref{subsubsec:blazarsc}, the smaller $\gamma$-ray--radio relation would result in a model that cannot by itself fit the data and would require a flaring component. In effect, this is, on average, equivalent to our choice of a single population of blazars with a given $\gamma$-ray--radio relation. We do not expect this choice to have a major impact on our results. However, as we will discuss in Section \ref{subsubsec:ques}, the blazar duty cycle is a remaining uncertainty, and it could impact predictions for the number of blazars that are observable by {\it Fermi}.
\item The model has been updated to account for effect of source confusion in the in blazar contribution to the EGB (see Sections \ref{subsubsec:sourceconf} and \ref{sec:results}). As will be shown in Section \ref{sec:results}, this has the effect of increasing the blazar background at lower energies.
\item The SID of FSRQs has been updated following the analysis of \citet{vp07} ($\Gamma_0 = 2.45$, $\sigma_0 = 0.15$) and correcting for spectral bias as in \citet{ven09} (see Section \ref{subsubsec:blazarSID}). Thus, the collective spectrum of blazars is not as hard as that presented in \citet{ste96} and does not exhibit as much curvature.
\end{enumerate}

\subsubsection{Remaining Questions}\label{subsubsec:ques}

There remain a few open questions, the answers to which will impact the determination of the blazar contribution to the EGB.  The blazar duty cycle, which dictates the amount of time a blazar spends in the quiescent state versus the flaring state, remains uncertain, as do questions of the amount the flux increases during flaring and whether the spectral index changes during flaring.  Analyses of EGRET blazar spectral indices found no evidence of systematic changes in spectral index with flaring \citep{nan07,vp07}, and {\it Fermi} observations of individual blazars have thus far revealed no systematic changes in spectral index with time or flux \citep{09pks1454-354,10pks1510-089,103C454.3}.  As such, we are justified in assuming that the blazar spectral index remains constant, on average, with time.  However, we acknowledge that the uncertainty of blazar variability parameters could have an impact on the counts of faint blazars and the $\gamma$-ray--radio correlation.  This uncertainty will decrease as more data from {\it Fermi} become available.

Another open question is that of the nature of blazar spectra over the entire {\it Fermi} energy range.  We treat blazar spectra as unbroken power laws over this range and in many observed blazars, this does appear to be a reasonable approximation.  However, in at least a few cases, {\it Fermi} has found evidence that blazar spectra can break \citep{lat09a,specprop10,sedlat10}.  Whether such observations are representative of the entire blazar population is presently unclear, as is the nature of the breaks.  In any case, spectral breaks are likely to impact the collective unresolved blazar spectrum mostly at the high end of the {\it Fermi} energy range. It is also possible that the spectra of harder blazars\footnote{We do not include a possible contribution from BL Lacs
for lack of a comparable radio data set.} will compensate for that of softer blazars at higher energies \citep{ven10a}.

\subsection{Star-forming Galaxies}\label{subsec:sfginput}

As discussed in Section \ref{subsec:galaxyform}, the Milky Way is a source of substantial \gray emission arising primarily the decay of $\pi^{0}$ meson produced in inelastic collisions of cosmic rays with interstellar gas.  Thus, it is expected that other star-forming galaxies emit \grays through the same interactions and that unresolved star-forming galaxies could provide a substantial contribution to the EGB.  Thus far, the {\it Fermi}-LAT Collaboration has reported detections of two nearby irregular galaxies \citep[the SMC and the LMC;][]{SMClat10, LMClat10}, two starburst galaxies \citep[M82 and NGC253;][]{starblat10}, and M31, a galaxy similar to our own \citep{M31lat10}.  As such, whatever the contribution to the EGB from unresolved star-forming galaxies, it will not have changed substantially in the {\it Fermi} data with respect to EGRET data as {\it Fermi} has resolved only a handful of star-forming galaxies.

The determination of the star-forming galaxy contribution relies on knowledge of the star formation rate of galaxies and their gas content, both of which are subject to substantial observational and theoretical uncertainties.  At relatively low redshifts ($z \la 1.5$), nebular and forbidden emission lines (e.g., H$\alpha$, O~II, and O~III) can be used to trace star formation in galaxies\footnote{For review of observational techniques of measuring star formation, see \citet{ken98}.}.  At higher redshifts ($z \sim 1-5$), the redshifted UV continuum is used.  However, both observational techniques are subject to uncertainties in dust extinction and the stellar initial mass function.  Alternatively, in noting that UV radiation from young stars is absorbed by interstellar dust and reradiated in the infrared, measurements of the far-infrared (FIR) continuum can be used to trace star formation.  Nevertheless, early-type galaxies can exhibit substantial FIR emission possibly due to dust heating by older stars or AGNs.  Furthermore, infrared measurements are hindered by emission from our own Galaxy.  Given these factors,  the large degree of scatter present in the measurements (about a factor of a few; \citealp{leb09,ly10}) of the cosmic SFR density is not surprising.  

The gas content of galaxies is even more uncertain, particularly at high redshift.  The amount of  ${\rm H_2}$ in a galaxy is determined from measurements of CO emission, while HI is determined from measurements of the $21$-cm line.  However, the CO-to-${\rm H_2}$ conversion varies depending on the metallicity and radiation field of a given region and the opacity of the molecular clouds containing CO. The $21$-cm surveys, on the other hand, extend only out to $z \sim 0.05$.  At higher redshifts, measurements of the HI density of the universe rely on damped Lyman-$\alpha$ absorbers observed in the Lyman-$\alpha$ forest of quasar spectra, but the nature of these systems is still the subject of much debate \citep[see {\it e.g.},][]{kul10,per10}.  Furthermore, the connection between the total gas and star formation rate is complex~\citep{put09}.  While it is fairly well established that stars form in GMCs and hence that star formation traces ${\rm H_2}$, the amount of HI varies between galaxies and does not appear to be correlated with star formation \citep{big08,ler08}. From both the observational and the theoretical points of view, the transition from HI to ${\rm H_2}$ and the formation of GMCs remain uncertain \citep{ler08}.

Given the uncertainty surrounding key elements of the determination of the star-forming galaxy contribution to the EGB, our approach does not focus on a particular model.  Instead, we employ several families of models that rely on separate sets of assumptions each with advantages and caveats.  Using this approach we seek to explore various possibilities for the star-forming galaxy contribution and highlight the uncertainty in such a calculation.  The strategies we employ are summarized as follows:
\begin{enumerate}
\item Relate the galaxy gas mass to its stellar mass assuming a gas fraction that evolves with redshift.
\item Relate the galaxy \gray luminosity to its SFR, which, in turn, is related to an observable for which there is a redshift distribution ({\it e.g.}, IR luminosity).
\item Relate the cosmic density of gas in star-forming galaxies to the star formation rate density.
\end{enumerate}

\subsubsection{The Schechter Function Model}\label{subsubsection:inputschecter}

In this approach, we relate the galaxy gas mass to its stellar mass assuming an evolving gas fraction (for details of the model, see Section \ref{subsubsection:schecter}).  We employ the Schechter parameters of the stellar mass functions as determined by \citet{els08} and the evolving gas fraction as determined by \citet{pap10}.  Since extensive spectroscopic surveys are, as yet, unavailable, \citet{els08} make use of combined data from the multi-band photometry of the GOODS-MUSIC catalog and the {\it Spitzer Space Telescope}.  In so doing, they infer the stellar masses of galaxies from photometric data by fitting the mass-to-light ratios of galaxies to stellar population templates.  Such a procedure is subject to a considerable degree of uncertainty, particularly arising from that of dust extinction\footnote{In order to mitigate the effect of the uncertainty in dust extinction, \citet{els08} make use of ${\rm K_S}$-band M/L ratios since dust absorption is small for longer wavelengths.} and the usage of photometric redshifts.  \citet{els08} estimate the mean uncertainty in their stellar mass estimates to be about a factor of two, though they did not estimate the possible uncertainty resulting from the usage of photometric redshifts.  

\citet{pap10} study the relationship between the star formation rate and stellar mass of high redshift galaxies selected at a constant comoving number density and derive a gas fraction that evolves\footnote{Note that we extrapolate the functional form of their gas fraction to low redshifts.} as $(1+z)^{0.9}$.  They estimate that the uncertainty on the gas mass is $\sim$ 0.11 dex.  

The advantage of this kind of model is that it is based on observations that will be continuously refined with time.  However, we note that the relationship between the stellar mass of a galaxy and its total gas content is unclear, though perhaps with better observations and a better theoretical understanding, the relationship will be better determined.

\subsubsection{The IR Luminosity Function Models}\label{subsubsec:inputIR}

\begin{figure}[t]
\begin{center}
\resizebox{2.5in}{!}
{\includegraphics[trim = 0 0 0 1mm, clip]{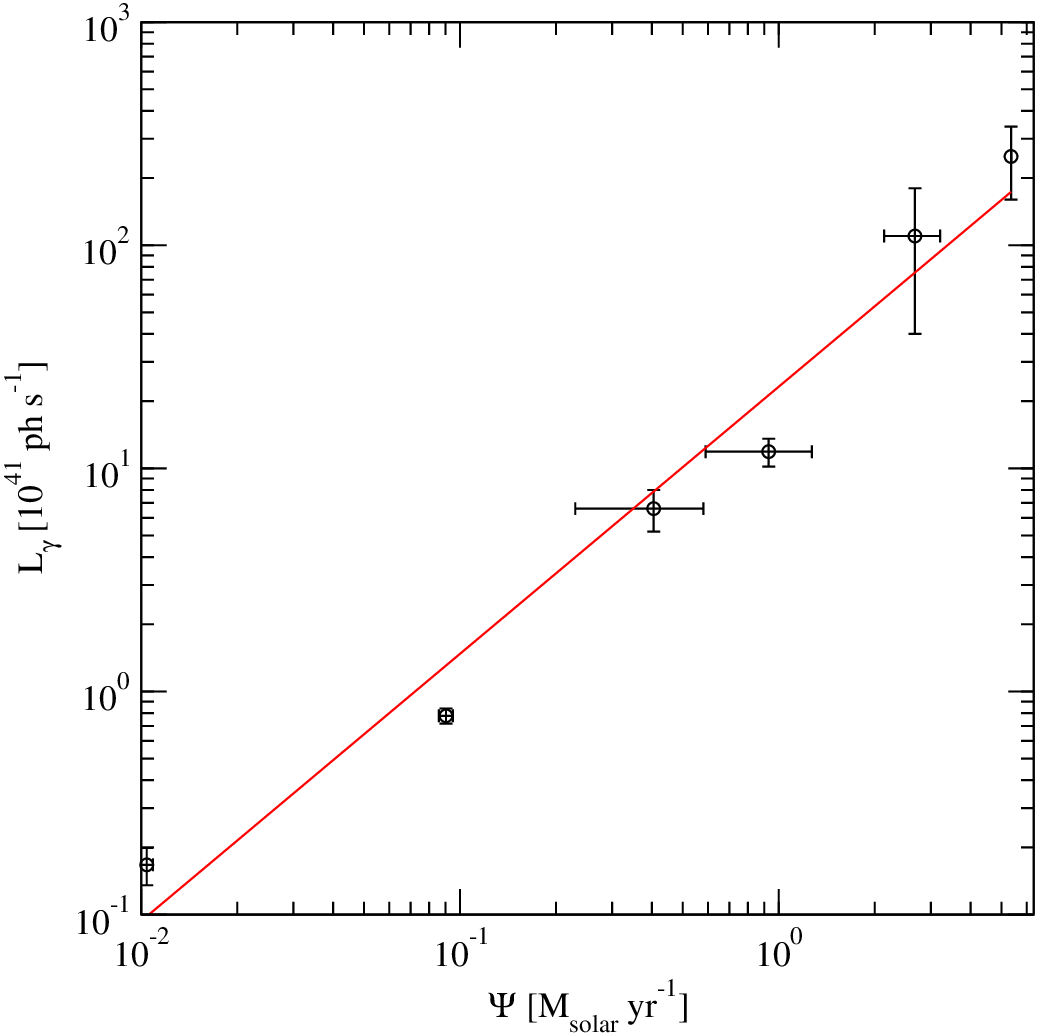}}
\caption{The \gray luminosities of Local Group galaxies plotted versus their star formation rates (data points; see Table \ref{table:LgammavsSFR}).  Also plotted is the power-law fit relating the \gray luminosity to the SFR (solid red line): $L_{\gamma} \propto \Psi^{1.2}$.
\label{fig:lgammasfr}}
\end{center}
\end{figure}

\begin{table}[t]
\caption{Observables for Local Group Galaxies}
\begin{center}
\begin{tabular}{|c c c|}
\hline
ID\footnote{\emph{SMC:} $L_{\gamma}$ from \citet{SMClat10}. SFR from \citet{law10} corrected for IMF. \emph{LMC:} $L_{\gamma}$ from \citet{LMClat10}. SFR from \citet{law10} corrected for IMF. \emph{M31:} $L_{\gamma}$ from \citet{M31lat10}. SFR from \citet{tab10} and \citet{wil03} corrected for IMF. \emph{MW:} $L_{\gamma}$ calculated for $\left<q_{\rm H}\right> \sim 1.4 \times 10^{-25} \, {\rm s^{-1} \, H^{-1}}$ and $M_{\rm H} \sim 7 \times 10^{9} M_{\odot}$ \citep{boi99}. SFR taken from \citet{rob10} corrected for IMF. \emph{NGC253:} $L_{\gamma}$ from \citet{starblat10}. SFR from IR measurements given by \citet{san03}. \emph{M82:} $L_{\gamma}$ from \citet{starblat10}. SFR from IR measurements given by \citet{san03}.  Note that the errors in the SFRs reflect only statistical uncertainties and assume that Equation \ref{eqn:IRSFRconv} is exactly correct.} & $L_{\gamma} \, ({\rm ph \, s^{-1}})$ & $\Psi \, (M_{\odot} \, {\rm yr^{-1}})$ \\ \hline
SMC & $(1.7 \pm 0.3) \times 10^{40}$ & $0.01 \pm (4.8 \times 10^{-4})$ \\
LMC & $(7.8 \pm 0.6) \times 10^{40}$ & $0.1 \pm (4.6 \times 10^{-3})$ \\
M31 & $(6.6 \pm 1.4) \times 10^{41}$ & $0.4 \pm 0.18$ \\
Milky Way & $(1.2 \pm 0.2) \times 10^{42}$ & $0.93 \pm 0.34$ \\
NGC253 & $(1.1 \pm 0.7) \times 10^{43}$ & $2.7 \pm 0.5$ \\
M82 & $(2.5 \pm 0.9) \times 10^{43}$ & $5.4 \pm (5.9 \times 10^{-4})$  \\
\hline
\end{tabular}
\end{center}
\label{table:LgammavsSFR}
\end{table}

In this approach, we assume that the \gray luminosity of a star-forming galaxy can be related to its star formation rate as a power law. In order to determine this relationship, we fit \gray luminosities of Local Group galaxies calculated from {\it Fermi} measurements\footnote{For the Milky Way, we calculate the \gray luminosity from $q_{\rm H}$ averaged over the whole galaxy (as discussed in Section \ref{subsec:galaxyform}) and the $M_{\rm H}$ taken from \citet{boi99}.} to their star formation rates either taken from the literature or calculated from IR measurements and converted to SFR via Equation \ref{eqn:IRSFRconv}.  In all cases, the SFR is calculated assuming (or corrected to) the IMF of \citet{cha03}.  We find the best-fit power law to be given by $L_{\gamma} \, [10^{41}{\rm ph \, s^{-1}}] \sim 24.0\times \Psi^{1.2}$. This fit is consistent with that obtained by the {\it Fermi} Collaboration, $L_{\gamma} \propto \Psi^{1.4 \pm 0.3}$ \citep{M31lat10}. Observables for Local Group galaxies used in this analysis are given in Table \ref{table:LgammavsSFR} and plotted with the fit in Figure \ref{fig:lgammasfr}. In a manner similar to that of blazars, we can relate the \gray luminosity of a star-forming galaxy to its total IR luminosity, convolve with an IR luminosity function, and integrate with respect to IR luminosity and redshift.

While the use of IR luminosity functions taken from observations is possible, it is difficult to deconvolve the contributions from obscured AGNs and mergers.  As such, we use the semi-empirical IR luminosity functions determined from the halo-occupation--based methodology of \citet{hop10} for both star-forming and starburst galaxies (in which enhanced star formation due to major mergers is taken into account)\footnote{Note that we do not consider any contribution from the so-called calorimetry effect for starburst galaxies as such an effect is likely to be small \citep{ste06}. We have assumed the same form of the $\pi^0$-decay spectrum for starburst galaxies as for star-forming galaxies.}. While the \citet{hop10} IR luminosity functions match available observations fairly well, we note that there is considerable debate over the roles of AGNs and mergers in driving star formation, the evolution of galaxies, and the determination of the IR luminosities of massive systems.  Furthermore, the $\gamma$-ray--SFR correlation is based on rather uncertain estimates of the SFRs and \gray luminosities of one normal galaxy (our own), two irregular galaxies, and two starburst galaxies, all of which could, in principle, exhibit different star formation properties.  As more data become available from {\it Fermi}, the $\gamma$-ray--SFR correlation will be further tested.  If it proves robust, then studies of the normal galaxy contribution the EGB could have implications for large-scale--structure formation and the evolution of galaxies with cosmic time.

\subsubsection{The Strong Coupling $\gamma$-ray--Star Formation Rate Model}\label{subsubsection:inputcsfr}

In this approach, we seek to relate the cosmic density of hydrogen in star-forming galaxies to the cosmic star formation rate.  Observations of nearby galaxies have indicated that localized star formation traces the density of ${\rm H_2}$ but there is no direct correlation with HI \citep{big08,ler08}. While in principle, there should be some relationship between the amount of HI, the amount of ${\rm H_2}$, and the SFR in a galaxy, other factors such as density fluctuations and turbulence make such a relationship complex. Therefore, while acknowledging that the relationship between HI and SFR is unclear, we simply assume that the amount of HI in star-forming galaxies is, \emph{on average}, comparable to that of ${\rm H_2}$ within the optical radius of the galaxy (see Section \ref{subsubsec:csfr}).  Hence, in this model, we take the \gray luminosity to be roughly proportional to the \emph{square} of the SFR.  We stress that the assumption that the star formation rate is proportional to the available gas density \emph{only applies to galaxies that are actively forming stars.} This assumption does not apply to galaxies at very high redshifts which may contain substantial amounts of gas but have yet to begin forming stars. However, we only include star-forming galaxies out to $z \sim 4$, so the higher redshift galaxies will not impact on our results. Given that the best-fit power-law index determined for Local Group galaxies in Section \ref{subsubsec:inputIR} is $\sim 1.2$ and given the proximity of the resulting unresolved spectrum to the {\it Fermi} measurements of the EGB (see Section \ref{sec:results}), we consider this model to be reflective of an upper limit to the star-forming galaxy contribution to the EGB. 

\subsection {The Fermi Spectrum and Unresolved Sources vs. Truly Diffuse Mechanisms}

The {\it Fermi} observations have placed significant constraints on extragalactic dark matter
annihilation \citep{cir09,dm10, ack10}. Currently, there is no evidence of quark-annihilation features and spectral lines seen in the EGB spectrum, features that would be a clear annihilation signal~\citep[see {\it e.g.},][]{ste89,rud91}. The observed spectrum does not match that expected from dark matter annihilation, placing constraints on any dark matter annihilation contribution to the EGB \citep{dm10}. Therefore, it is probable that dark matter annihilation $\gamma$-rays, if present, provide only a minor contribution to the EGB. 

The same argument about matching spectra can be made regarding the contribution from electromagnetic cascades produced by very high and ultrahigh energy cosmic-ray interactions as the resulting spectrum would be significantly harder than the observed spectrum  \citep{kal09, ber10, ahl10,ven10}.

\section{Results}\label{sec:results}

\begin{figure}[t]
\begin{center}
\resizebox{2.5in}{!}
{\includegraphics[trim = 0 0 0 0.5mm, clip]{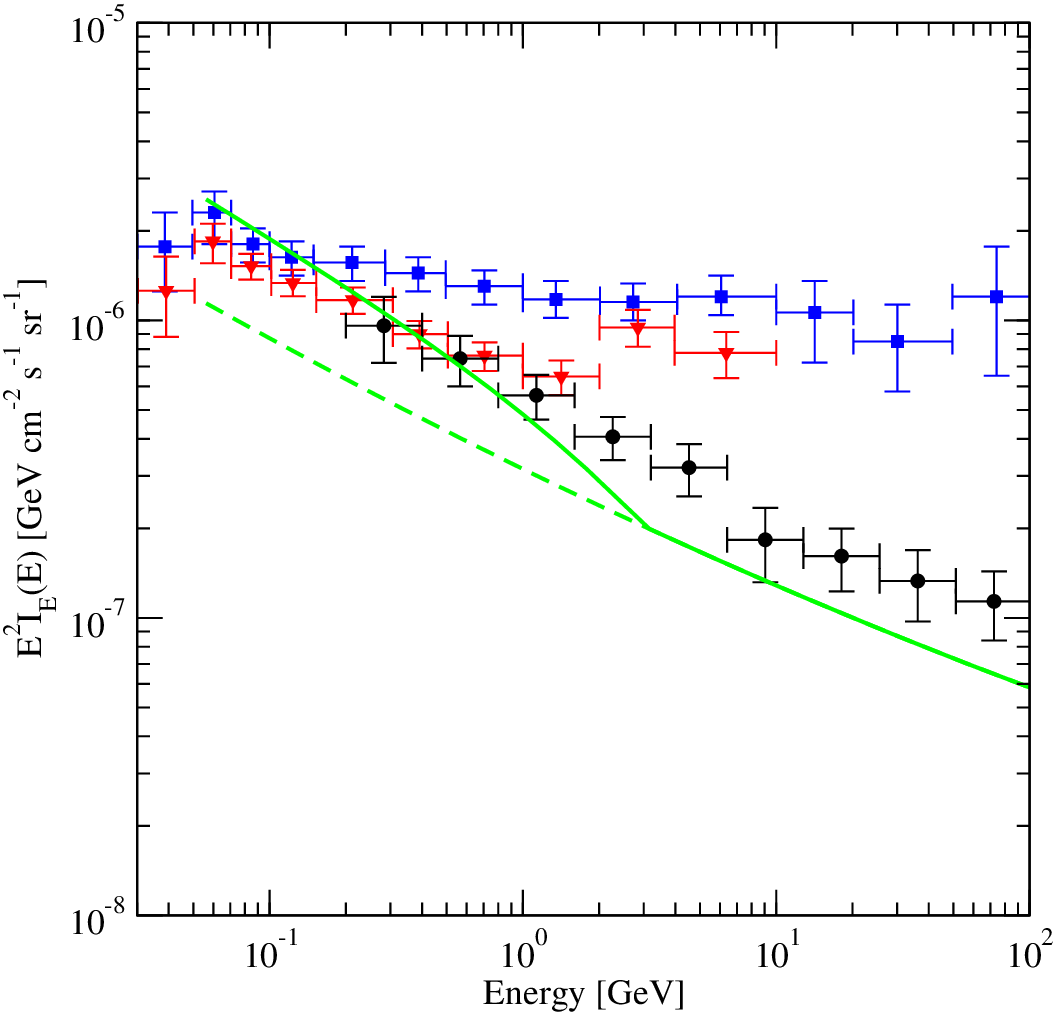}}
\caption{The collective spectrum of unresolved FSRQs. \emph{Solid green line:} The spectrum accounting for source confusion. \emph{Dashed green line:} The spectrum without accounting for source confusion. \emph{Black circles:} The {\it Fermi} measurement of the spectrum of the EGB as determined in \citet{lat10}. \emph{Blue squares:} The EGRET measurement of the spectrum of the EGB as determined by \citet{sre98} and confirmed by the analysis of \citet{shk08} and S. D. Hunter (private communication). \emph{Red triangles:} The EGRET measurement of the spectrum of the EGB as determined by \citet{str04}.
\label{fig:egrbfsrq}}
\end{center}
\end{figure}

The calculated spectrum of the unresolved FSRQ contribution to the EGB (see Sections \ref{subsec:blazarform} and \ref{subsec:inputblazars}) is plotted in Figure \ref{fig:egrbfsrq}. For comparison, we include the {\it Fermi} analysis of the EGB \citep{lat10}, two analyses\footnote{The two sets of EGRET data points result from two different estimations of the galactic foreground emission.} of the EGRET EGB \citep{sre98,str04}, and the calculation of the collective spectrum of unresolved FSRQs ignoring the effect of source confusion. Our results clearly show that the effect of source confusion is to reduce the number of resolved sources, increasing the collective intensity of unresolved blazars, particularly below $\sim 1$ GeV energy. Thus, accounting for source confusion modifies the predicted spectrum such that the EGRET and {\it Fermi} measurements of the EGB below $\sim 1$ GeV are both compatible with unresolved FSRQs. In contrast, the better angular resolution of the {\it Fermi}-LAT above $\sim$ 1 GeV allows it to resolve more blazars resulting in a limiting flux that is dominated by the {\it Fermi}-LAT sensitivity rather than source confusion. Thus, the collective spectrum of FSRQs breaks at $\sim 3$ GeV\footnote{The actual break should be more gradual since in our calculations we used the approximate broken angular resolution curve shown in Figure \ref{fig:angres}.}. At energies above $\sim 1$ GeV, the predicted collective spectrum of FSRQs falls below the data points, though they are likely consistent with the data within the uncertainties in the galactic foreground emission model.  Note also that the collective FSRQ spectrum exhibits much less curvature than seen in \citet{ste96}. This is because the spread in the SID in our current model is much smaller than that of the \citet{ste96} model.

\begin{figure}[t]
\begin{center}
\resizebox{2.5in}{!}
{\includegraphics[trim = 0 0 0 0.5mm, clip]{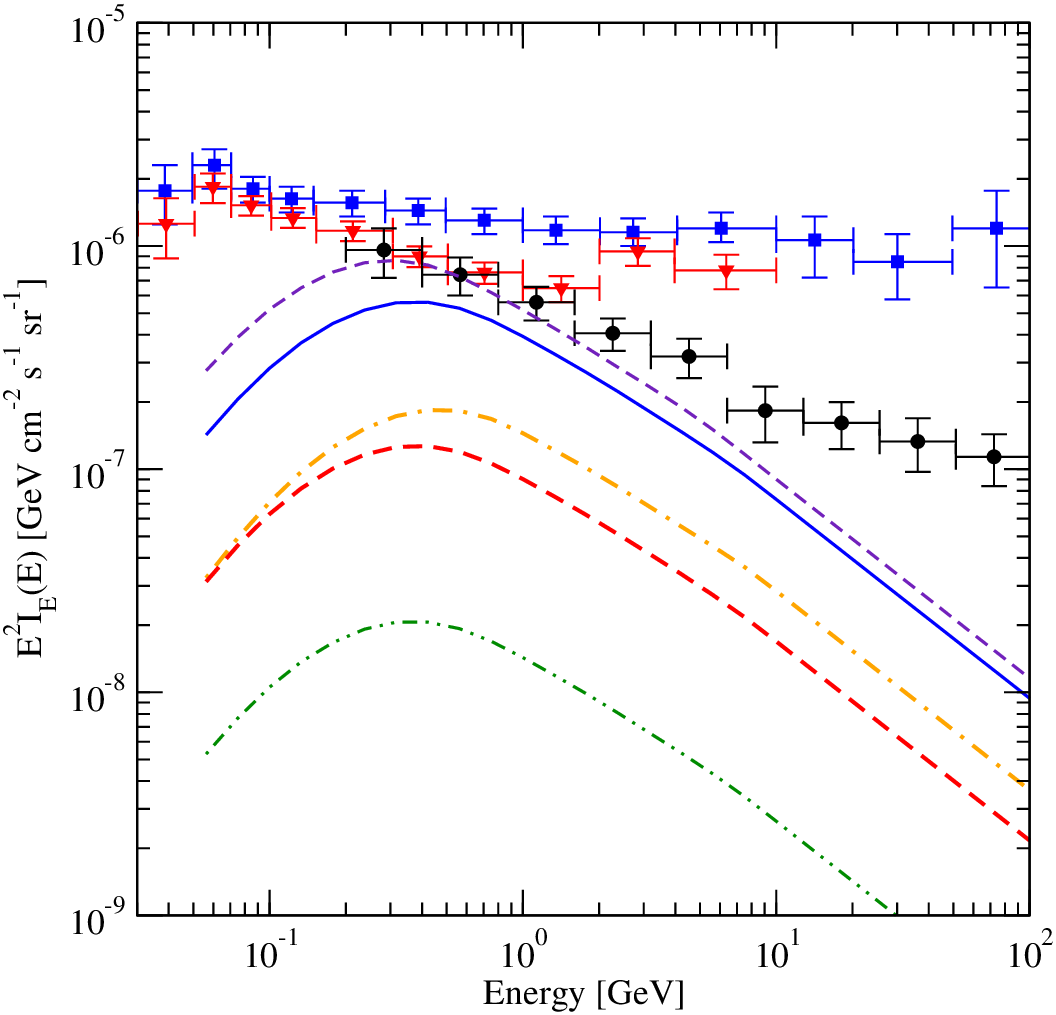}}
\caption{The collective spectrum of unresolved star-forming galaxies. \emph{Dashed indigo line:} The spectrum determined from the strong coupling model (see Sections \ref{subsubsec:csfr} and \ref{subsubsection:inputcsfr}. \emph{Solid blue line:} The spectrum determined from the IR luminosity function model (see Sections \ref{subsubsec:infrared} and \ref{subsubsec:inputIR}). \emph{Dot-dashed yellow line:} The spectrum determined from the IR luminosity function model assuming no gas evolution. \emph{Dashed red line:} The spectrum determined from the Schechter function model (see Sections \ref{subsubsection:schecter} and \ref{subsubsection:inputschecter}). \emph{Double dot-dashed line:} The spectrum of the starburst contribution alone determined from the IR luminosity function model.
\label{fig:egrbsfgal}}
\end{center}
\end{figure}

In Figure \ref{fig:egrbsfgal}, we plot the spectra of the unresolved star-forming galaxy contributions to the EGB calculated for the models discussed in Sections \ref{subsec:galaxyform} and \ref{subsec:sfginput}.  For comparison, we include the spectrum of the unresolved starburst galaxy contribution alone that we determined from the best-fit IR luminosity function of \citet{hop10}. For the spectrum of starburst galaxies, we have assumed the same form of the $\pi^0$ decay spectrum as for star-forming galaxies. The range in the calculations of the overall contribution to the EGB from unresolved star-forming galaxies spans about an order of magnitude indicating the degree of uncertainty in such a calculation\footnote{Though, we note that each individual model is subject to its own uncertainty. As such, the degree of uncertainty is likely even more than an order of magnitude. Within the range of our various predictions of the EGB from star forming galaxies, we agree with the results of the model of~\cite{fie10} and \citet{mak10}.}

We note that even though our most extreme model could possibly explain the lowest energy {\it Fermi} data points (and possibly, within systematics, a couple others), it cannot explain the EGRET data points below $300$ MeV. The \citet{str04} EGRET data points (minus the two highest energy data points) with the {\it Fermi} data points resemble a \emph{featureless} power law, while the spectra of unresolved star-forming galaxies do not.  Notably, the data points show no indication of a $\pi^0$-decay ``bump'' at the energies at which the contribution of the star-forming galaxies should peak. 

\section{Discussion \& Conclusions}

We have calculated the spectral shape of the contribution of unresolved FSRQs to the EGB assuming that the \gray luminosity of an FSRQ is, on average, proportional to its radio luminosity \citep{gir10, lat10,ghi10a,mah10}, and also accounting for the effects of source confusion.  We have demonstrated that the combination of the source density predicted by the \citet{dun90} FSRQ radio luminosity function and the strong energy dependence of the {\it Fermi}-LAT angular resolution \emph{increases} the contribution of unresolved FSRQs to the EGB at energies below $1$ GeV. The resulting overall spectrum predicted by the fit to the {\it Fermi} source count distribution reproduces well the spectrum of the EGRET and {\it Fermi} EGB measurements below $1$ GeV, but falls below the data points above $1$ GeV.  We have also calculated the spectral shape of the contribution of unresolved star-forming galaxies to the EGB for several relations for the \gray luminosity of a star-forming galaxy.  We find that, depending on the model, the overall amount of the contribution of star-forming galaxies to the EGB may be more or less significant, though regardless of the model considered, the spectrum of unresolved star-forming galaxies is unable to explain the combined spectrum of the low-energy EGRET EGB measurements and the {\it Fermi} EGB measurements. Similar calculations for starburst galaxies alone indicate that they account for at most about $1$\% of the EGB, in agreement with the conclusion reached by \citet{ste06}.

The similarity of the collective spectrum of unresolved FSRQs to the combined spectrum of the EGRET and {\it Fermi} EGB measurements, as demonstrated by our results, is striking.  In fact, we note that as predicted in \citet{ste99}, the inclusion of the effect of source confusion in the calculation could provide an explanation for the similarity between the EGRET and {\it Fermi} EGB measurements at energies of hundreds of MeV.  The density of FSRQs predicted by the model is sufficiently large such that at these energies, {\it Fermi} would not be able to resolve many more FSRQs than EGRET did, and the FSRQ contribution to the EGB would remain the same for {\it Fermi} as for EGRET.  Thus, if unresolved FSRQs \emph{do} comprise the bulk of the EGB emission, then one would expect such similarity between the EGRET and {\it Fermi} measurements at these energies.  At energies above $1$ GeV, the {\it Fermi}-LAT angular resolution improves substantially with respect to that of EGRET, and as such, {\it Fermi} would be able to resolve more blazars at higher energies than EGRET could, resulting in a decrease in the {\it Fermi} EGB with respect to the EGRET EGB, an effect which is possibly indicated by comparing the EGRET and {\it Fermi} results\footnote{A caveat is that the uncertainties in the subtraction of the galactic foreground emission at the higher energies are considerable owing to the uncertainty in the distributions of both gas and cosmic rays in the Galaxy. Furthermore, the instrumental backgrounds of EGRET and the {\it Fermi}-LAT are different, so it is difficult to make a direct comparison between the two. We should also note that in our calculations, we have neglected the population of BL Lacs, which, due to their hard spectra, are likely to have more of a contribution at energies above $\sim 10$ GeV. Notably, {\it Fermi} has resolved as many BL Lacs as FSRQs.}.

In contrast, no such high-energy separation between the EGRET EGB and the {\it Fermi} EGB is predicted for star-forming galaxies as all but the closest are too faint to be resolvable by {\it Fermi}. Furthermore, we note that the EGRET EGB measurements provide no indication of a turnover in the spectrum as would be expected if unresolved star-forming galaxies comprise the bulk of the EGB emission. Rather, the spectrum of unresolved star-forming galaxies is \emph{inconsistent} with the combined spectrum of the EGRET and {\it Fermi} EGB measurements\footnote{As previously noted, the {\it Fermi}-LAT was designed to reach its optimal effective area for \grays with energies near and above $\sim 1$ GeV, whereas EGRET was designed to reach its optimal effective area for \grays with energies near and above $\sim 100$ MeV.  Also, the {\it Fermi}-LAT detector has a significantly higher instrumental background at $100$ MeV than EGRET did (S. D. Hunter, private communication). Thus, the EGB was not reported by {\it Fermi} for energies below $200$ MeV \citep{lat10}.}. We also note that the lack of a turnover in the EGRET data is not simply the result of systematics (S. D. Hunter, private communication), since the uncertainties in all of the galactic foreground models used to determine the EGB from the EGRET and {\it Fermi} data are quite small at these energies. Finally, we note that at energies above $\sim 1$ GeV, the spectrum of unresolved star-forming galaxies is steeper than the spectra of the EGB data.  As such, we conclude that however significant the contribution of star-forming galaxies to the EGB may be, it is not sufficient to explain the EGB\footnote{The effect of Compton interactions mentioned in Section \ref{subsec:galaxyform} does not alter this conclusion as it only modifies the spectrum above 10 GeV for normal galaxies \citep{str10} and the starburst galaxy contribution to the EGB is negligible (See Figure \ref{fig:egrbsfgal}.)}.

Within the range of our various predictions of the EGB from star forming galaxies, we agree with the results of the models of both~\cite{fie10} (which suggests that star-forming galaxies may comprise the bulk of the EGB) and \citet{mak10} (which suggests that star-forming galaxies can account for less than $10\%$ of the EGB). This underscores the range of uncertainty in the calculation for star-forming galaxies\footnote{One noteworthy difference between our model and that of \citet{fie10} is that in order to relate the gas mass of a galaxy to its star formation rate, \citet{fie10} makes use of the Schmidt-Kennicutt relation. However, in doing so, they estimate the disk sizes of galaxies to high redshifts. Given that the uncertainties in these quantities are likely to be considerable (and given the uncertainty already present in the calculation), we considered alternative approaches. Nevertheless, we note that the Schmidt-Kennicutt relation was included in the inputs to both the IR luminosity models and the Schechter function model. In the IR luminosity model, we tested the impact of changing the Schmidt-Kennicutt law by performing the calculation for the \citet{hop10} IR luminosity function calculated using a steeper Schmidt-Kennicutt relation and found that it had very little impact on our results.}.

The featureless spectrum of the EGB deduced by {\it Fermi} is intriguing when one considers the possibility of features that could arise from phenomena such as breaks in blazar spectra, absorption of high-energy \grays from unresolved blazars, \gray emission from unresolved star-forming galaxies, \gray emission from dark matter annihilation, and \grays from electromagnetic cascades initiated by very high and ultrahigh energy particle interactions with the extragalactic background light.  The spectra of these potential contributions to the EGB differ considerably from that of the FSRQs \citep{sil84,ste85,rud88,ste89, ste89a,rud91,ull02,ando07,kal09,sie09,ber10,ahl10,ven10}. However, recent {\it Fermi} observations have placed significant constraints on dark matter annihilation \citep{cir09,dm10, ack10}, and presently there is no clear evidence of annihilation features above the background continuum.  As such, it appears that any putative contribution to the EGB from dark matter annihilation is relatively minor. The possible contribution to the EGB from electromagnetic cascades is constrained by the relative steepness of the EGB spectrum, though cascades could play a role at higher energies \citep{kal09,ber10,ahl10,ven10}. An apparent explanation for the featureless power-law spectrum of the 
EGB as presently deduced could be that unresolved blazars provide the dominant contribution to the EGB, given that their collective spectrum is roughly consistent with that of the EGB.

Therefore, we conclude that, contrary to the result given by~\citet{pop10}, the {\it Fermi} observations do not rule out the possibility that the EGB is dominated by emission from unresolved blazars.

\section*{Acknowledgments}

We thank David Thompson, Stan Hunter, and Olaf Reimer for discussions of the EGRET
detector characteristics and EGRET data. We thank Marco Ajello for sending
the results of his Monte Carlo simulations of the {\it Fermi}-LAT efficiency {\it vs.} source
flux. We also thank Dawn Erb for comments regarding the infrared surveys of 
star-forming galaxies at high redshifts and Matt Malkan, Vasiliki Pavlidou, Jane Rigby, and the anonymous referee for helpful discussions. T.M.V. acknowledges support by an appointment to the NASA Postdoctoral Program at the Goddard Space Flight Center, administered by Oak Ridge Associated Universities through a contract with NASA.

\clearpage

\bibliography{ms_bibtex}
\bibliographystyle{apj}

\end{document}